\documentclass[aps,prb,twocolumn,showpacs,superscriptaddress]{revtex4}
\usepackage{epsfig}
\usepackage{graphicx}
\usepackage{amsmath}
\usepackage{amsfonts}
\usepackage{amssymb}
\usepackage{bm}

\begin{document}
\title{Numerical study of critical properties and
hidden orders in dimerized spin ladders}
\author{Sandra J. Gibson}
\affiliation{Department of Physics, Laurentian University, Sudbury,
ON, P3E 2C6 Canada}
\author{R. Meyer}
\affiliation{Department of
Physics, Laurentian University, Sudbury, ON, P3E 2C6 Canada}
\affiliation{Department of Mathematics and Computer Science,
Laurentian University, Sudbury, ON, P3E 2C6 Canada }
\author{Gennady Y. Chitov}
\affiliation{Department of Physics, Laurentian University,
Sudbury, ON, P3E 2C6 Canada}
\date{\today}
%
%xxxxxxxxxxxxxxxxxxxxxxxxxxxxxxxxxxxxxxxxxxxxxxxxxxxxxxxxxxxxxxxxxxxxxxxxxxxxxx
%
\begin{abstract} Dimerized antiferromagnetic spin-$\frac12$ ladders
are known to exhibit a quantum critical phase transition in the ground
state, the existence or absence of which is dependent on the
dimerization pattern of the ladder. The gapped phases cannot be
distinguished by the conventional Landau long-range order
parameter. However, they possess a non-local (hidden) string order
parameter, which is non-zero in one phase and vanishes in the
other. We use an exact diagonalization technique to calculate ground
state energies, energy gaps and string order parameters of dimerized
two- and three-leg Heisenberg ladders, as well as a critical scaling
analysis to yield estimates of the critical exponents $\nu$ and $\beta$.
\end{abstract}
\maketitle
%
%xxxxxxxxxxxxxxxxxxxxxxxxxxxxxxxxxxxxxxxxxxxxxxxxxxxxxxxxxxxxxxxxxxxxxxxxxxxxxx
%
\section{Introduction}
%
%xxxxxxxxxxxxxxxxxxxxxxxxxxxxxxxxxxxxxxxxxxxxxxxxxxxxxxxxxxxxxxxxxxxxxxxxxxxxxx
%
Spin ladders have been the focus of significant theoretical interest
in the past decade or more. This work has been grounded by the experimental
discovery of several spin ladder and spin-Peierls compounds exhibiting
various fascinating phenomena.\cite{Dagotto96}
Many of the important features of these materials have long been understood (For reviews see e.g.
Refs.~\onlinecite{Dagotto96,Giamarchi04}). However, significant discoveries may
yet be made as crucial questions about the underlying nature of these systems remain.

One of the most peculiar properties of spin ladders is that the existence of a gap
(i.e. mass) depends on the number of legs. The spin excitations in a
$m$-leg spin ladder are gapped if $m$ is even, and the system is
gapless (quantum critical) when the number of legs $m$ is odd. The
even-$m$-leg ladders are interesting examples of spin liquids, which
are currently under enormous scrutiny. The spin liquid has, by
definition, a gap which is not due to long-range order or apparent
symmetry breaking. This is hard to understand in terms of the basic postulates
of the conventional Landau paradigm for phase transitions.\cite{Landau}

The purpose of this study is to explore the question of the existence of
``hidden orders'' associated with the massive quantum phases of
dimerized Heisenberg ladders with antiferromagnetic couplings. It is
known that single dimerized Heisenberg spin-$\frac12$ chains are gapped,
and that when two chains are coupled into a ladder, the
system is gapped even without dimerization.\cite{Giamarchi04} Counterintuitively, the
dimerized two- and three-leg ladders can also be gapless. The critical (gapless)
lines of the dimerized two- and three-leg ladders were first
conjectured in Ref.~\onlinecite{Delgado96} and have been confirmed by
subsequent numerical and analytical work.
\cite{Delgado98,Kotov99,Cabra99,Nersesyan00,Okamoto03,Nakamura03,Almeida,Chitov08}
The characterization of the unconventional quantum phases in such
systems in terms of more exotic types of orders may ultimately help to uncover novel physics which lies beyond the Landau paradigm.

We present numerical results on finite size spin-$\frac12$ systems with total
number of spins ($N$) of up to $N = 30$. These results are obtained
here using an exact diagonalization (ED) technique based on the
Lanczos algorithm.\cite{Lanczos} The total available computer memory is the
most important factor limiting the maximum system sizes accessible using this method.
This is because the dimension of the Hamiltonian matrix grows exponentially fast
(as $2^N$ for a spin-$\frac12$ state). Although restricted to relatively small system
sizes, the ED method is considered an "unbiased method" useful for calculating almost
any physical quantity. Our implementation of the Lanczos algorithm uses periodic boundary conditions and includes an
explicit restart step with greater than $5$ restarts to mitigate loss of
orthogonality in the Lanczos basis.
%
%xxxxxxxxxxxxxxxxxxxxxxxxxxxxxxxxxxxxxxxxxxxxxxxxxxxxxxxxxxxxxxxxxxxxxxxxxxxxxx
%
\section{Two-leg Ladder}
\begin{figure}[hb!]
\epsfig{file=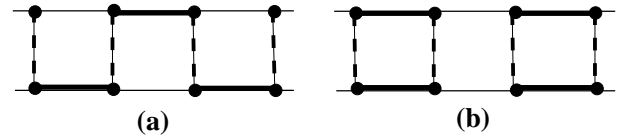,width=0.45\textwidth,angle=0} \caption{Dimerized
two-leg ladder. Bold/thin/dashed lines represent the stronger/weaker
chain coupling $J(1 \pm \delta)$ and rung coupling $J_\bot$,
respectively.  Dimerization patterns: (a) - staggered; (b)-
columnar.} \label{Dim_Lad}
\end{figure}

%
%xxxxxxxxxxxxxxxxxxxxxxxxxxxxxxxxxxxxxxxxxxxxxxxxxxxxxxxxxxxxxxxxxxxxxxxxxxxxxx
%
We begin by presenting the results for a two-legged ladder with intrinsic
dimerization. A similar program is carried out for the three-leg ladder,
detailed in Sec. III. In line with the previous work\cite{Chitov08,Khalada07}
which used the bond-mean-field theory,\cite{Azzouz03} we have considered the
two possible dimerization patterns of the ladder (staggered and columnar) on an
equal footing (cf. Fig.~\ref{Dim_Lad}).
%
%xxxxxxxxxxxxxxxxxxxxxxxxxxxxxxxxxxxxxxxxxxxxxxxxxxxxxxxxxxxxxxxxxxxxxxxxxxxxxx
%xxxxxxxxxxxxxxxxxxxxxxxxxxxxxxxxxxxxxxxxxxxxxxxxxxxxxxxxxxxxxxxxxxxxxxxxxxxxxxx
%

\pagebreak

The Hamiltonian of the dimerized spin ladder with $m$-legs is given by:
\begin{eqnarray}\label{Ham}
H &=& \sum_{n=1}^{N}\sum_{\alpha=1}^{m} J_{\alpha}(n)
\mathbf{S}_{\alpha}(n) \cdot \mathbf{S}_{\alpha}(n+1) \nonumber \\
&+& J_\bot \sum_{n=1}^{N}\sum_{\alpha=1}^{m-1} \mathbf{S}_{\alpha}(n) \cdot \mathbf{S}_{\alpha+1}(n),
\end{eqnarray}
where the dimerization occurs along the chains ($\alpha=1 \dots m$) only,
with the rung coupling $J_\bot$ taken as constant. All the spin exchange couplings
are antiferromagnetic. The dimerization patterns are then defined as:
\begin{eqnarray}
J_{\alpha}(n)=J[1+(-1)^{n+\alpha}\delta] & \mbox{(staggered)} \label{staggered}\\
J_{\alpha}(n)=J[1+(-1)^{n}\delta] & \mbox{(columnar)} \label{columnar}
\end{eqnarray}
with periodic boundary conditions along the chains and open boundary conditions
in the rung directions.
%xxxxxxxxxxxxxxxxxxxxxxxxxxxxxxxxxxxxxxxxxxxxxxxxxxxxxxxxxxxxxxxxxxxxxxxxxxxxxx
%
\subsection{Ground State Energies and Gaps}
%
%xxxxxxxxxxxxxxxxxxxxxxxxxxxxxxxxxxxxxxxxxxxxxxxxxxxxxxxxxxxxxxxxxxxxxxxxxxxxxx
%
%xxxxxxxxxxxxxxxxxxxxxxxxxxxxxxxxxxxxxxxxxxxxxxxxxxxxxxxxxxxxxxxxxxxxxxxxxxxxxx
\begin{figure}[ht!]
\includegraphics[width=8.0cm]{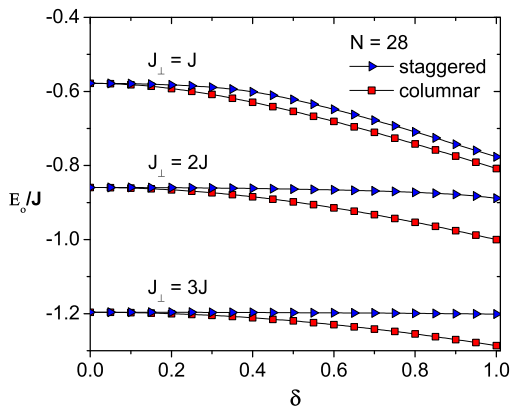}
\caption{(Color online) The ground state energies for the staggered and columnar
two-leg ladder calculated by exact diagonalization for $N = 28$.\label{gs_energies}}
\end{figure}
%xxxxxxxxxxxxxxxxxxxxxxxxxxxxxxxxxxxxxxxxxxxxxxxxxxxxxxxxxxxxxxxxxxxxxxxxxxxxxx
%
We find that the ladders with the columnar dimerization order always have the
lower ground state energy, and that they are always gapped. The results for
the ground state energies of the ladders with $N = 24$ are shown in
Figure~\ref{gs_energies} are very close to earlier
predictions for these systems from the mean-field theory.\cite{Chitov08}

\begin{figure}[ht!]
\includegraphics[width=8.0cm]{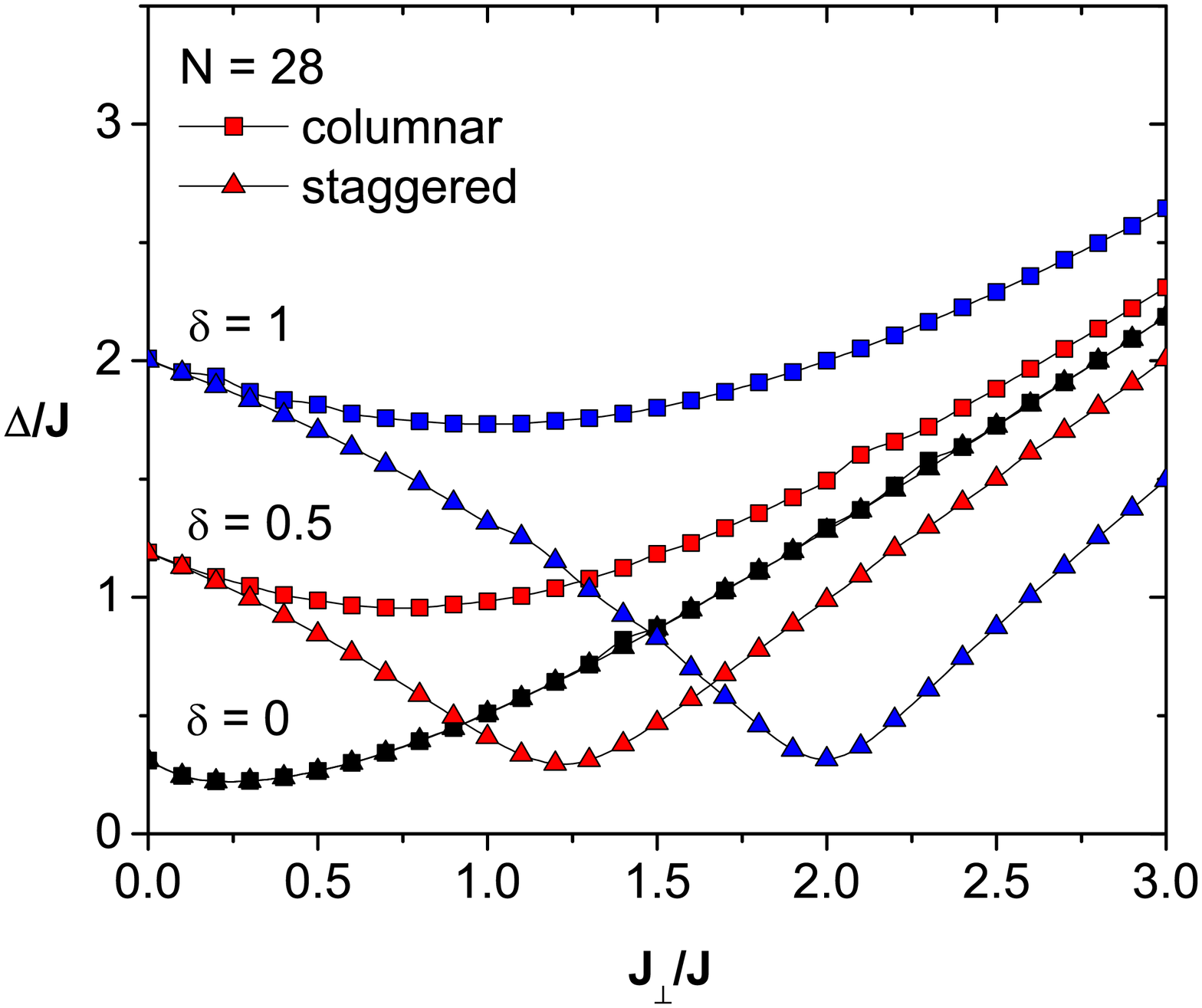}
\caption{(Color online) The ground state energy gap for the staggered and columnar
two-leg ladder calculated by exact diagonalization for $N = 28$.\label{gap_2leg}}
\end{figure}

Also in agreement with the earlier mean-field predictions,\cite{Chitov08} our ED
calculations confirm that the existence or absence of the quantum critical
phase transition between gapped phases depends on the dimerization pattern.
The \textit{columnar} arrangement of the strong and weak couplings on the
legs is never critical, whereas the \textit{staggered} configuration
possesses a quantum critical point. The characteristic results for the
singlet-triplet gaps are shown in Fig.~\ref{gap_2leg}.
%
%xxxxxxxxxxxxxxxxxxxxxxxxxxxxxxxxxxxxxxxxxxxxxxxxxxxxxxxxxxxxxxxxxxxxxxxxxxxxxx
%xxxxxxxxxxxxxxxxxxxxxxxxxxxxxxxxxxxxxxxxxxxxxxxxxxxxxxxxxxxxxxxxxxxxxxxxxxxxxx
%

\begin{figure}[ht!]
\includegraphics[width=7.0cm]{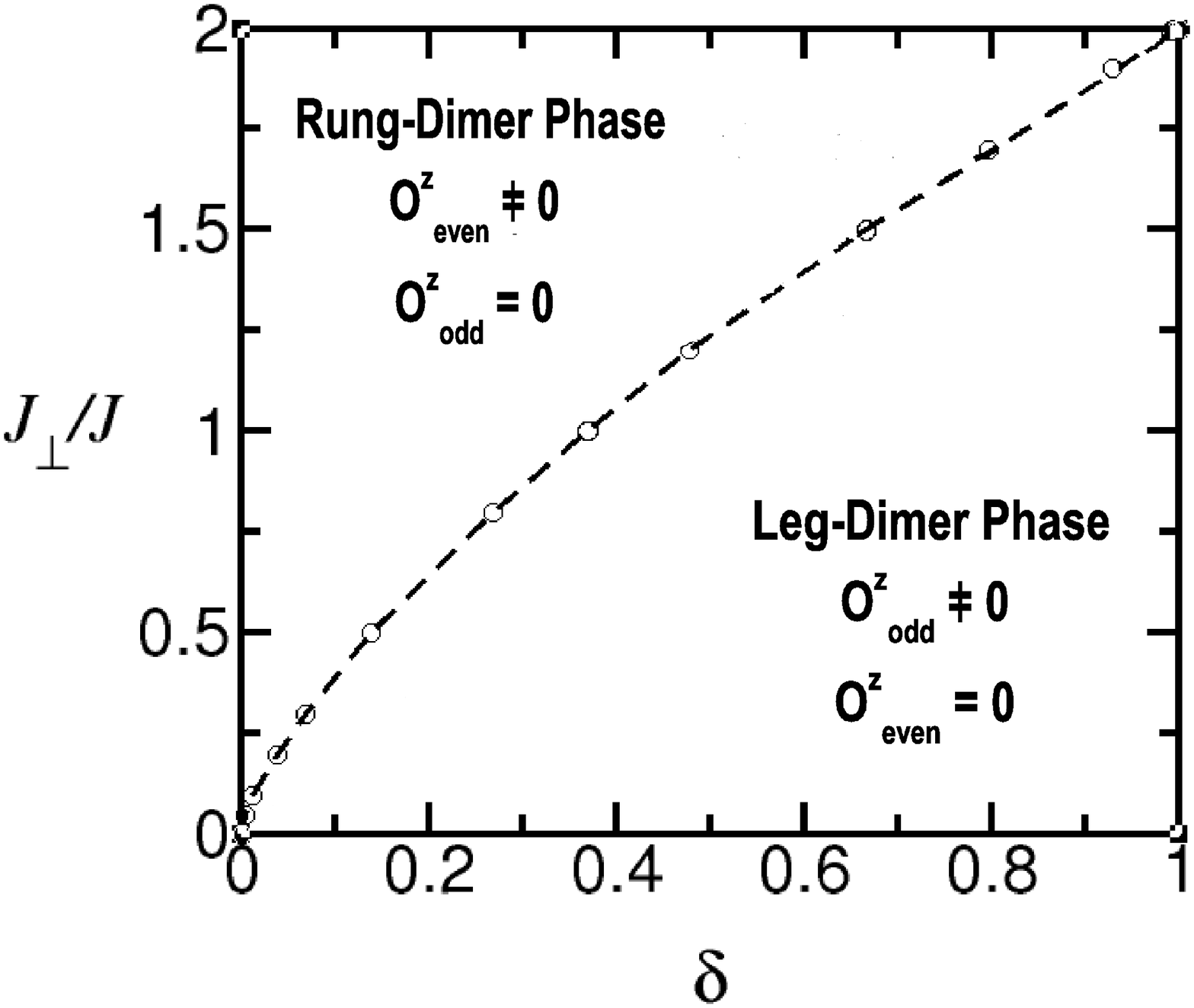}
\caption{Two-leg ladder; critical line $J_{\perp c}(\delta)$ where the
gap of the staggered phase vanishes. Adapted from
Ref.~\onlinecite{Chitov08}, original data from
Ref.~\onlinecite{Okamoto03}.
\label{Crit2L}}
\end{figure}

\begin{figure*}[]\hfil
\includegraphics[width=8.0cm]{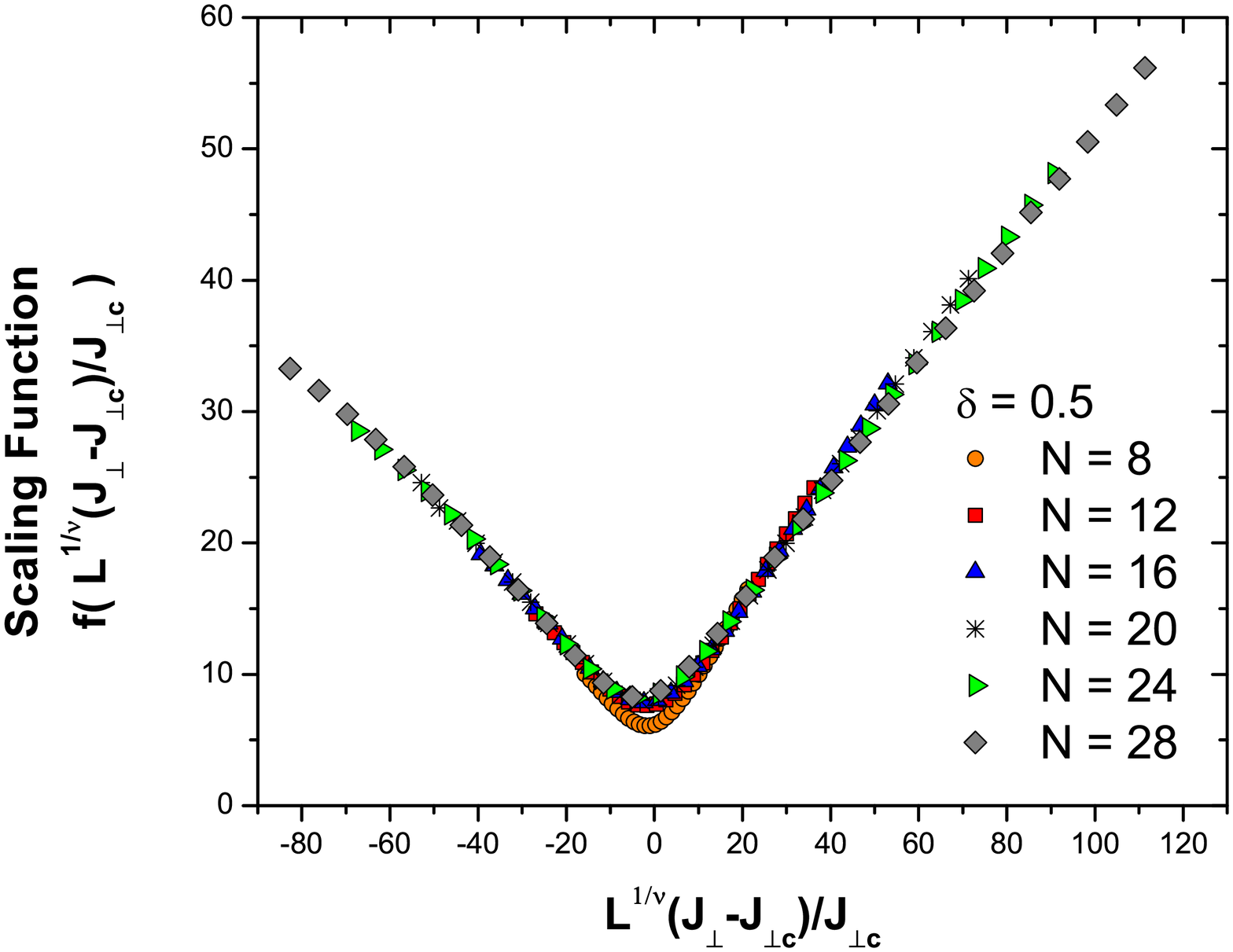}\hfil
\includegraphics[width=8.0cm]{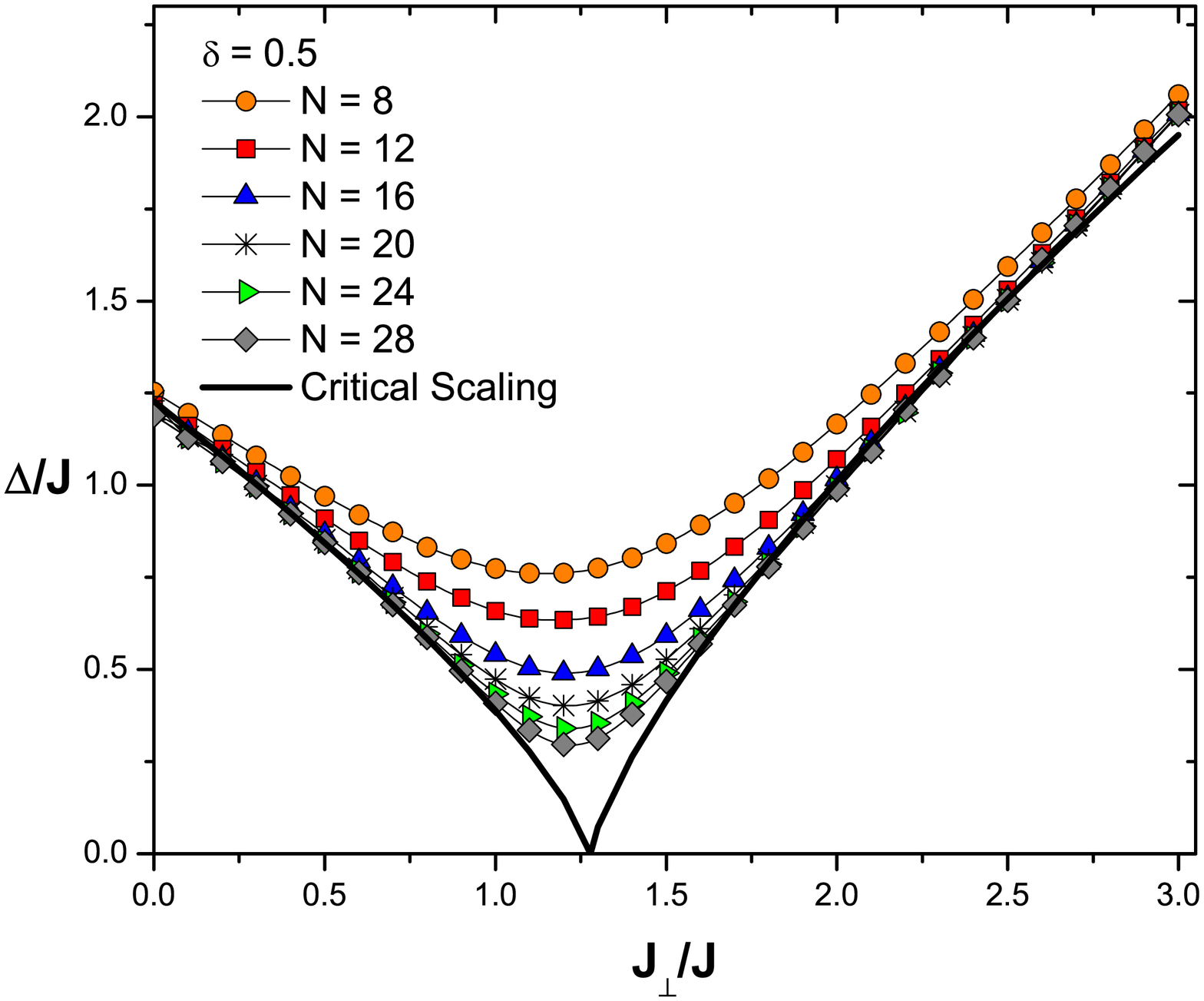}\hfil
\caption{(Color online) The collapsed scaling function (left panel)
and gap (right panel) for the staggered two-leg ladder with $\delta =
0.5$, calculated for $N = 8$ through 28 along with the fit using
optimized parameters $\nu \approx 0.755$ and $J_{\perp c} \approx
1.27$.\label{2l_gap_stag_delta05_sequence_fit} }
\end{figure*}

To take into account the finite size effects in the gaps
obtained by ED and use this data to extract the critical indices near
the transition point, we have applied a finite size scaling analysis.
\cite{Cardy} If we denote the critical coupling where the gap vanishes
as $J_{\perp c}$ (the phase diagram of the dimerized two-leg ladder is
shown in Fig.~\ref{Crit2L}) and the reduced coupling
as $g \equiv \frac{J_{\perp}-J_{\perp c}}{J_{\perp c}}$, the gap $\Delta_L(g)$
in the system of the linear dimension $L$ can be expressed as:
\begin{equation}
\label{scaling1} \Delta_{L}\left(g\right) =
L^{-1}f\left(gL^{\frac{1}{\nu}}\right)~,
\end{equation}
where $f(gL^{\frac{1}{\nu}})$ is a dimensionless
scaling function and $\nu$ is the critical index of the correlation
length. The unknown scaling function $f$ is determined by the best
collapse of the finite size ED data. The fitting gives $ \nu \approx 0.75$
(cf. Fig.~\ref{2l_gap_stag_delta05_sequence_fit}).

%
%xxxxxxxxxxxxxxxxxxxxxxxxxxxxxxxxxxxxxxxxxxxxxxxxxxxxxxxxxxxxxxxxxxxxxxxxxxxxxx
%xxxxxxxxxxxxxxxxxxxxxxxxxxxxxxxxxxxxxxxxxxxxxxxxxxxxxxxxxxxxxxxxxxxxxxxxxxxxxx
%
%xxxxxxxxxxxxxxxxxxxxxxxxxxxxxxxxxxxxxxxxxxxxxxxxxxxxxxxxxxxxxxxxxxxxxxxxxxxxxx
%xxxxxxxxxxxxxxxxxxxxxxxxxxxxxxxxxxxxxxxxxxxxxxxxxxxxxxxxxxxxxxxxxxxxxxxxxxxxxx
%
%xxxxxxxxxxxxxxxxxxxxxxxxxxxxxxxxxxxxxxxxxxxxxxxxxxxxxxxxxxxxxxxxxxxxxxxxxxxxxx
%
\subsection{String Order Parameter}
%
%xxxxxxxxxxxxxxxxxxxxxxxxxxxxxxxxxxxxxxxxxxxxxxxxxxxxxxxxxxxxxxxxxxxxxxxxxxxxxx
%
In the absence of the conventional Landau order parameter, the different gapped
phases in the ladder can only be distinguished by a non-local string order parameter
(SOP), which is non-zero in one phase and vanishes in the other. The SOP was
first introduced in 1989 by den Nijs and Rommelse,\cite{Rommelse89} who related it
with the gapped or Haldane phase in the spin-1 Heisenberg chain. In that system
there is no local symmetry breaking and therefore no local order parameter can be
identified. However, the SOP detects the hidden $Z_{2} \otimes Z_{2}$ symmetry
breaking \cite{Oshikawa92} in the Haldane phase. The SOP notion was later generalized for
two-leg spin ladders as:
\cite{Nishiyama,Shelton96,Kim00,Kim08,Fath01,White,Watanabe}
\begin{widetext}
\begin{equation}
\label{stringorder} O_\mathrm{odd/even}^{z} \equiv - \lim_{ \left|n-m\right|
\to \infty} \Big\langle S_\mathrm{rung/diag}^z(n) \exp \Big\{
i\pi\sum_{l=n+1}^{m-1} S_\mathrm{rung/diag}^z(l) \Big\} S_\mathrm{rung/diag}^z(m)
\Big\rangle~,
\end{equation}
\end{widetext}
where $S_\mathrm{rung}^z(n) \equiv S_1^z(n)+ S_2^z(n)$ and
$S_\mathrm{diag}^z(n) \equiv S_1^z(n)+ S_2^z(n+1)$.
%
%xxxxxxxxxxxxxxxxxxxxxxxxxxxxxxxxxxxxxxxxxxxxxxxxxxxxxxxxxxxxxxxxxxxxxxxxxxxxxx
\begin{figure}
\includegraphics[width=8.0cm]{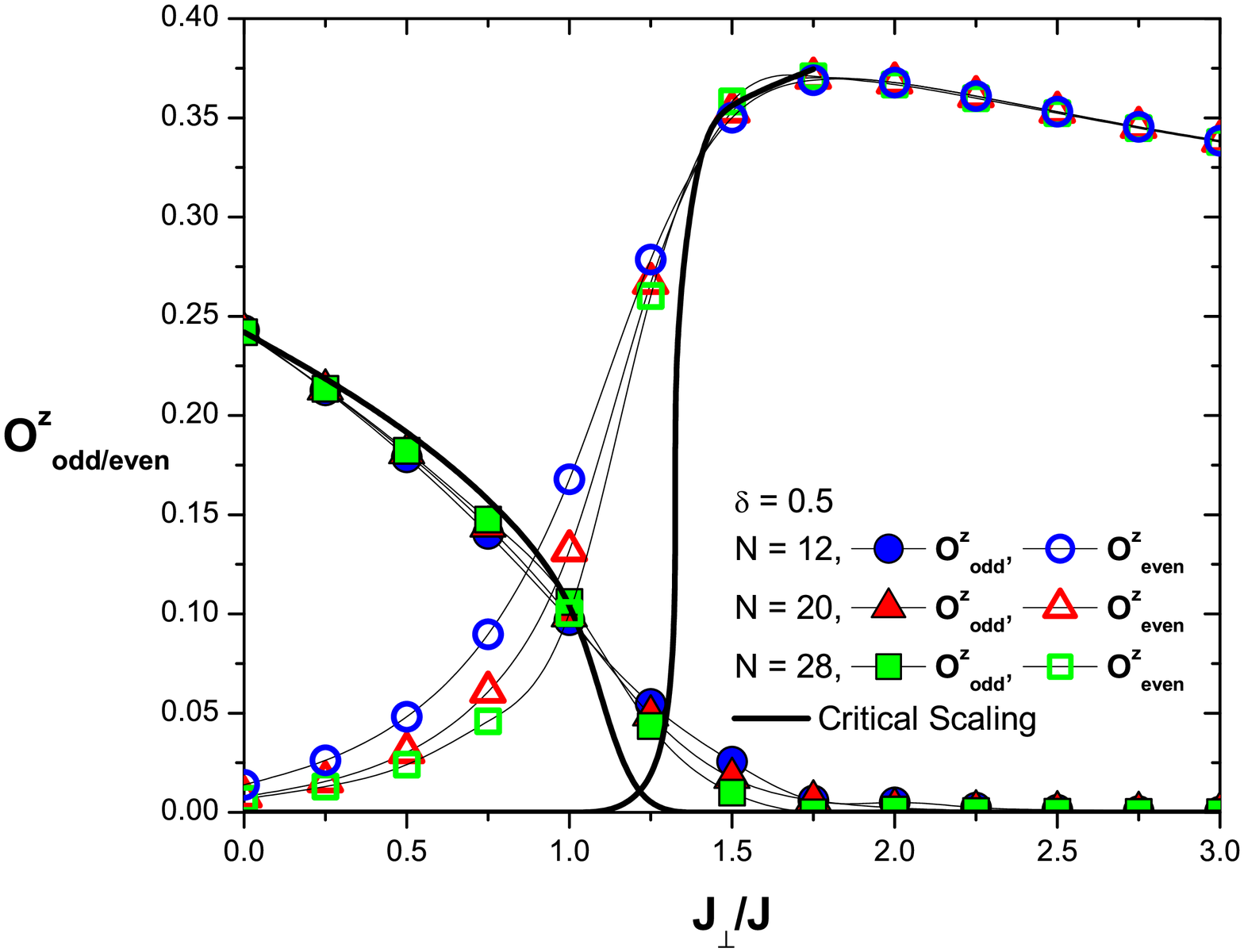}\hfil
\caption{(Color online) The odd and even string order parameters of
the staggered two-leg ladder with $\delta =0.5$. The ED data for
ladders with number of spins N = 12 to 28 are shown, along with the
critical scaling curves. \label{SOP_2leg_stag} }
\end{figure}
%xxxxxxxxxxxxxxxxxxxxxxxxxxxxxxxxxxxxxxxxxxxxxxxxxxxxxxxxxxxxxxxxxxxxxxxxxxxxxx
%
The even and odd SOP have been shown to be mutually exclusive in several numerical
and analytical studies.\cite{Fath01,Almeida} Previous numerical studies
\cite{Almeida, Nishiyama} have assumed a staggered dimerization, while here the SOP
were calculated for both configurations by the ED method. In agreement
with the previous studies, we find that the SOPs are featureless for the columnar
phase: in the whole phase space domain $(J_\perp, \delta)$ the even
string order parameter is non-zero, while the odd string order parameter
vanishes.

We find for the staggered phase that the even/odd SOP is non-zero only in the
rung-dimer/leg-dimer phases, respectively, and vanishes at the critical
point. The calculated SOPs are shown in Fig.~\ref{SOP_2leg_stag}.
%
%xxxxxxxxxxxxxxxxxxxxxxxxxxxxxxxxxxxxxxxxxxxxxxxxxxxxxxxxxxxxxxxxxxxxxxxxxxxxxx
\begin{figure*}
\includegraphics[width=8.0cm]{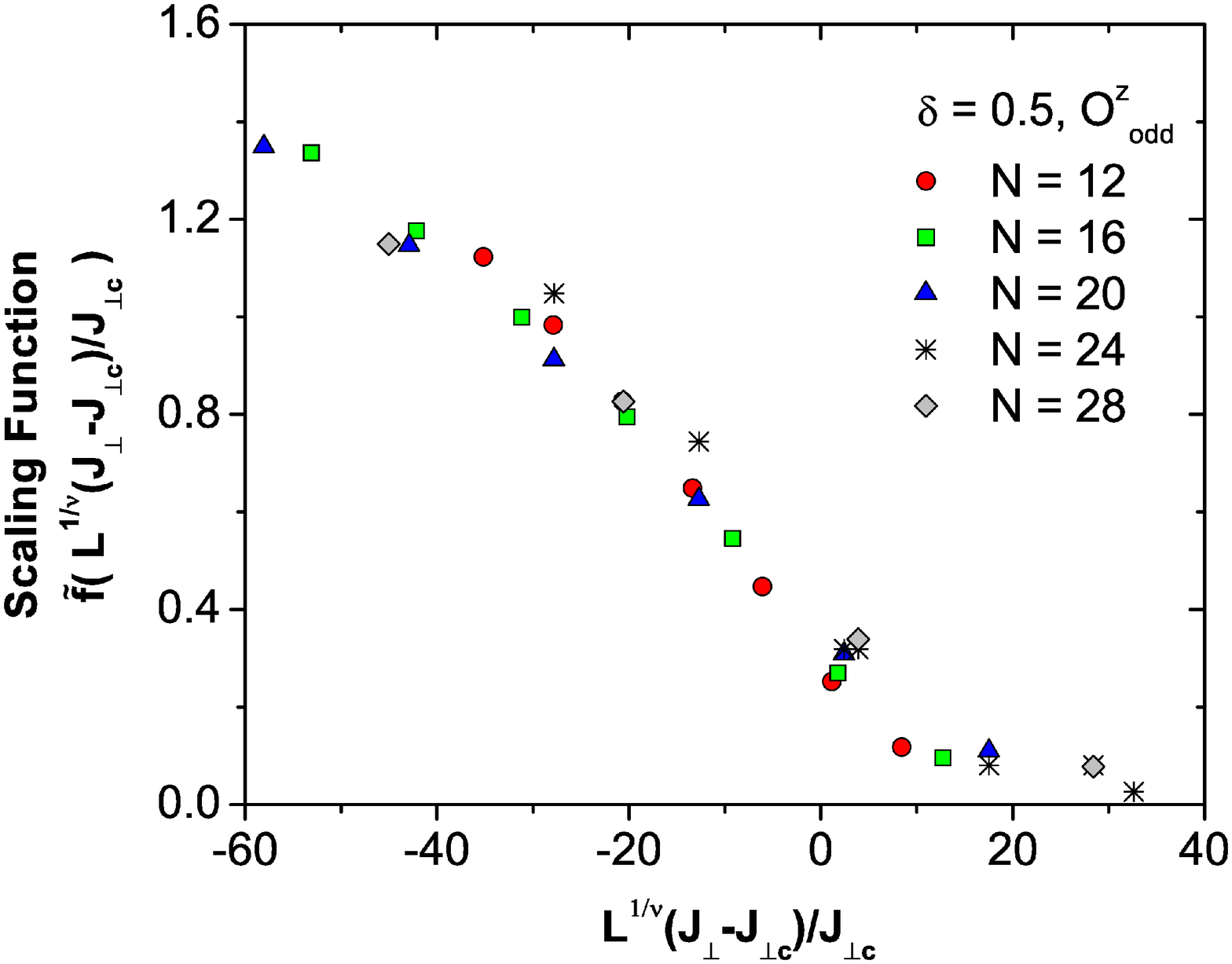}
\includegraphics[width=8.0cm]{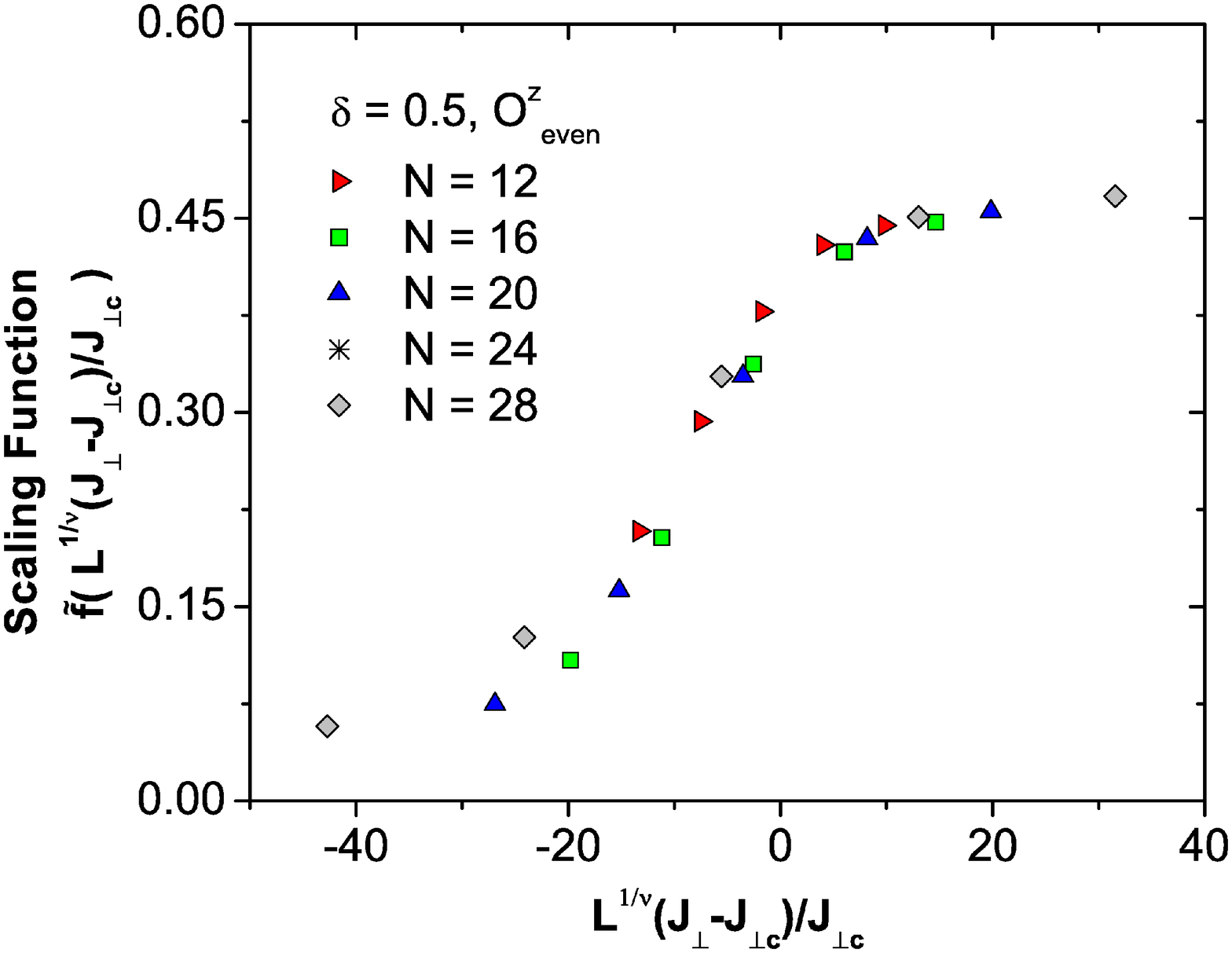}
\caption{(Color online) The lefthand (righthand) panel shows the fitted
scaling function for the odd (even) string order parameter of the
staggered two-leg ladder with $\delta = 0.5$. The data collapse for the
optimized scaling parameters $\nu = 0.7$, $J_{\perp c} = 1.21$, and
$\beta_\mathrm{odd} = 0.25$ ($\nu = 0.74$, $J_{\perp c} = 1.32$, and
$\beta_\mathrm{even} = 0.025$). \label{scalingfx_2leg_sop_collapse_even}}
\end{figure*}
%xxxxxxxxxxxxxxxxxxxxxxxxxxxxxxxxxxxxxxxxxxxxxxxxxxxxxxxxxxxxxxxxxxxxxxxxxxxxxx
%
The SOP vanishes at the critical point as $O_{\sharp}^{z} \propto g^{2
\beta}$ where $\beta$ is the critical index
of the order parameter.\cite{Hida92,Shelton96} Then the SOP in the system
of length $L$ scales as
\begin{equation}
\label{SOPscaling1} O^{z}_{L}(g) =
L^{-\frac{2\beta}{\nu}}\tilde{f}\left(gL^{\frac{1}{\nu}}\right)~,
\end{equation}
where the scaling function $\tilde{f}$ is determined as for the gap by the
requirement to collapse the numerical data for the systems of
different sizes onto the same curve. From the analysis of the ED data
for $\delta = 0.5$ we obtain a critical index $\beta_\mathrm{odd} \approx 0.25$ for
the odd string order and $\beta_\mathrm{even} \approx 0.025$ for the even string
order. The collapse of the scaling functions are shown for $\delta =
0.5$ in Fig.~\ref{scalingfx_2leg_sop_collapse_even}.
The same critical scaling analysis was performed for several values of $\delta$, and these
results are tabulated in Table~\ref{allindices}.
%
%xxxxxxxxxxxxxxxxxxxxxxxxxxxxxxxxxxxxxxxxxxxxxxxxxxxxxxxxxxxxxxxxxxxxxxxxxxxxxx
\begin{table}
\begin{center}
\caption{Critical scaling indices of the two-leg ladder.
\label{allindices} }
\begin{tabular}{|ccccc|} \hline \hline
$\delta$ & even/odd & $\beta$ & $\nu$ & $J_{\perp c}/J$\\ \hline
\hline 0.25 & even & 0.05294 & 0.758 & 0.775\\ 0.25 & odd & 0.285 &
0.787 & 0.759\\ \hline 0.5 & even & 0.0251 & 0.726 & 1.32\\ 0.5 & odd
& 0.215 & 0.698 & 1.21\\ \hline 0.75 & even & 0.0687 & 0.764 & 1.62\\
0.75 & odd & 0.235 & 0.648 & 1.64\\ \hline 1.0 & even & 0.0124 &
0.7306 & 2.08\\ 1.0 & odd & 0.227 & 0.7013 & 1.98\\ \hline
\end{tabular}
\end{center}
\end{table}
%xxxxxxxxxxxxxxxxxxxxxxxxxxxxxxxxxxxxxxxxxxxxxxxxxxxxxxxxxxxxxxxxxxxxxxxxxxxxxx
%
%xxxxxxxxxxxxxxxxxxxxxxxxxxxxxxxxxxxxxxxxxxxxxxxxxxxxxxxxxxxxxxxxxxxxxxxxxxxxxx
%
\section{Three-leg Ladder}
\begin{figure}[]
\epsfig{file=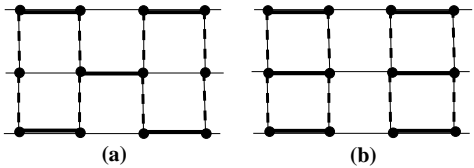,width=0.35\textwidth,angle=0} \caption{Dimerized
three-leg ladder. Bold/thin/dashed lines represent the stronger/weaker
chain coupling $J(1 \pm \delta)$ and rung coupling $J_\bot$,
respectively.  Dimerization patterns: (a) - staggered; (b)-
columnar.} \label{3lDim_Lad}
\end{figure}

%
%xxxxxxxxxxxxxxxxxxxxxxxxxxxxxxxxxxxxxxxxxxxxxxxxxxxxxxxxxxxxxxxxxxxxxxxxxxxxxx
%
Similar to the two-legged case the dimerized three-leg ladder occurs in two distinct
configurations, the staggered and columnar patterns (cf. Fig.~\ref{3lDim_Lad}).
The Hamiltonian of the intrinsically dimerized three-leg ladder is given by
Eq.~\eqref{Ham} with $m = 3$ and like the two-legged ladder the staggered/columnar
patterns are given by the coupling $J_{\alpha}(n)$ defined by
Eqs.~\eqref{staggered}/\eqref{columnar} respectively.
%
%xxxxxxxxxxxxxxxxxxxxxxxxxxxxxxxxxxxxxxxxxxxxxxxxxxxxxxxxxxxxxxxxxxxxxxxxxxxxxx
%xxxxxxxxxxxxxxxxxxxxxxxxxxxxxxxxxxxxxxxxxxxxxxxxxxxxxxxxxxxxxxxxxxxxxxxxxxxxxx
%
%xxxxxxxxxxxxxxxxxxxxxxxxxxxxxxxxxxxxxxxxxxxxxxxxxxxxxxxxxxxxxxxxxxxxxxxxxxxxxx
\begin{figure}[]
\includegraphics[width=8.0cm]{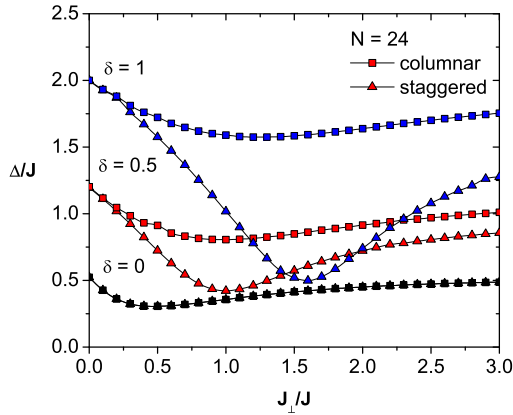}
\caption{(Color online) Energy gaps for the staggered and columnar
three-leg ladder dimerization patterns calculated by the exact
diagonalization for N = 24. \label{gap_3leg}}
\end{figure}
%xxxxxxxxxxxxxxxxxxxxxxxxxxxxxxxxxxxxxxxxxxxxxxxxxxxxxxxxxxxxxxxxxxxxxxxxxxxxxx
%
\subsection{Ground State Energies and Gaps}
\begin{figure}
\includegraphics[width=7.0cm]{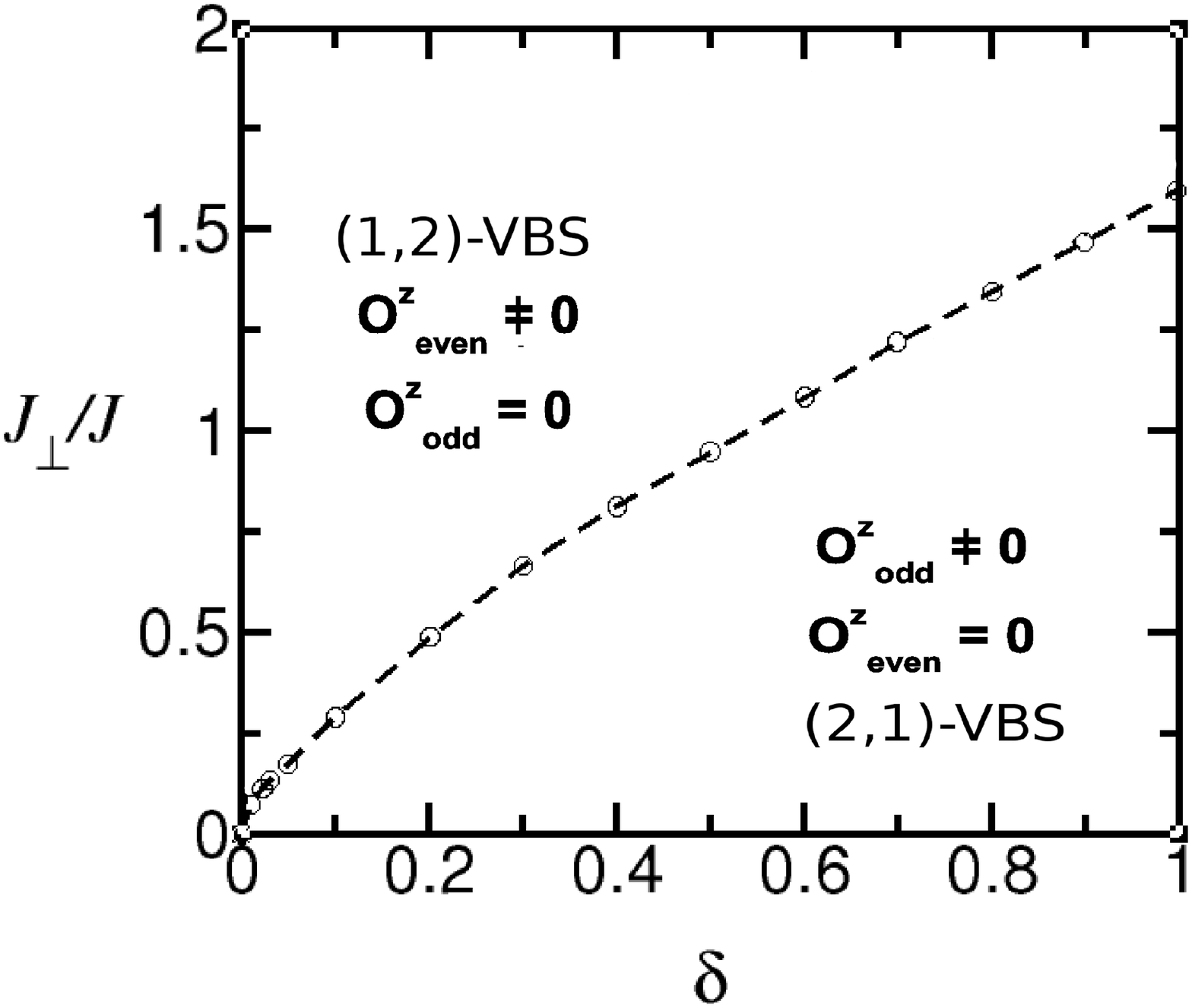}
\caption{Three-leg ladder; critical line $J_{\perp c}(\delta)$ where the
gap of the staggered phase vanishes. Adapted from Ref.~\onlinecite{Chitov08},
original data from Ref.~\onlinecite{Almeida}.
\label{Crit3L}}
\end{figure}

\begin{figure}
\includegraphics[width=8.0cm]{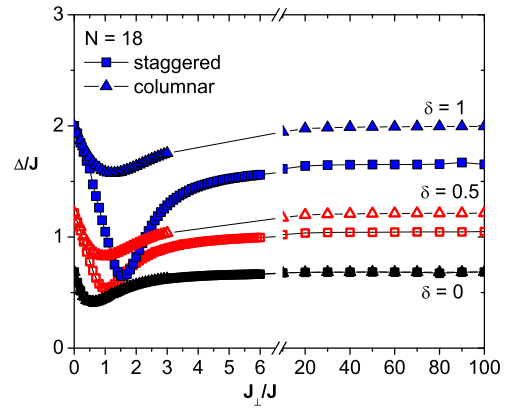}
\caption{(Color online) Three-leg ladder; gap saturation for strong rung couplings $J_{\perp}$.
\label{gap_3l_long}}
\end{figure}

\begin{figure*}\hfil
\includegraphics[width=8.0cm]{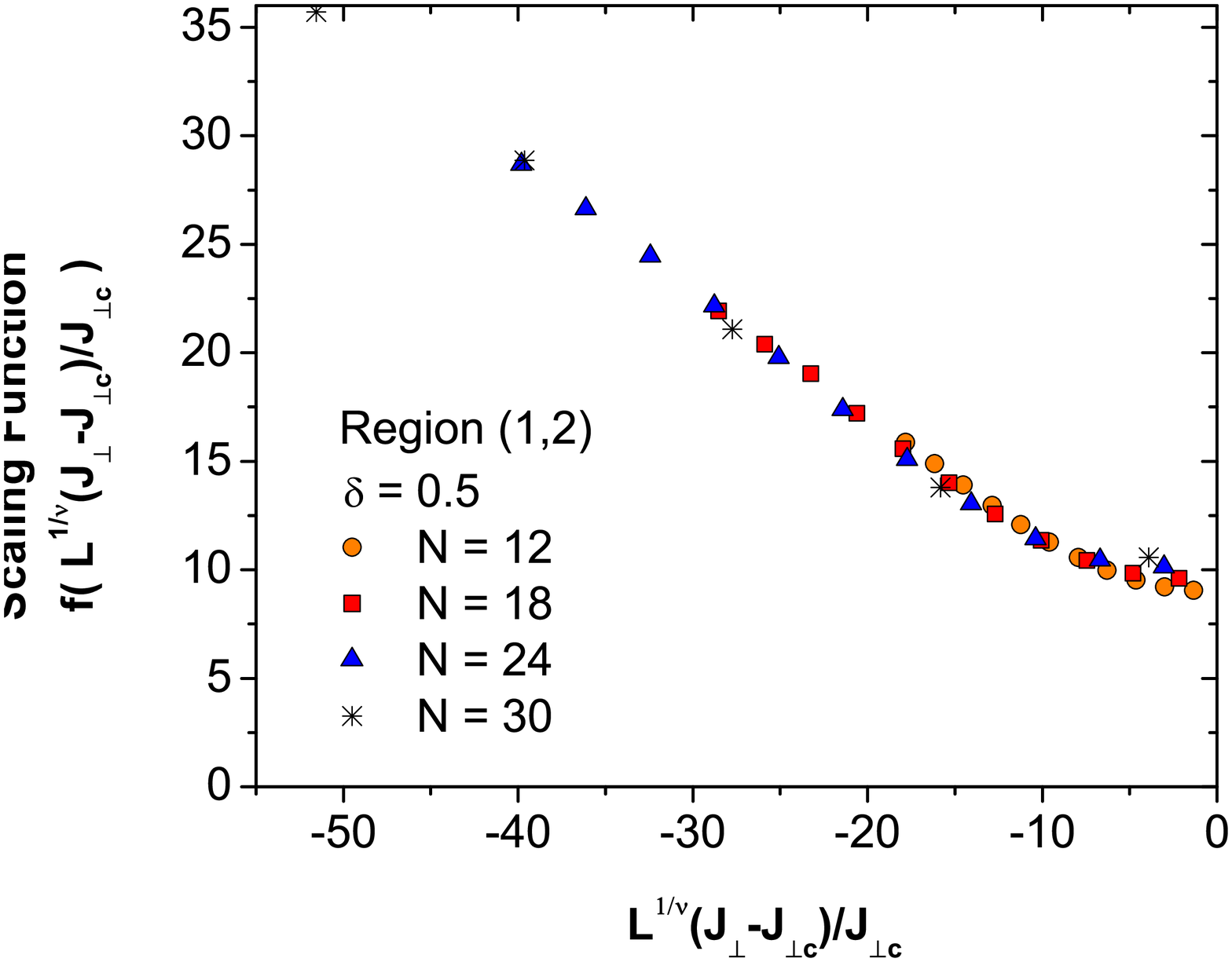}\hfil
\includegraphics[width=8.0cm]{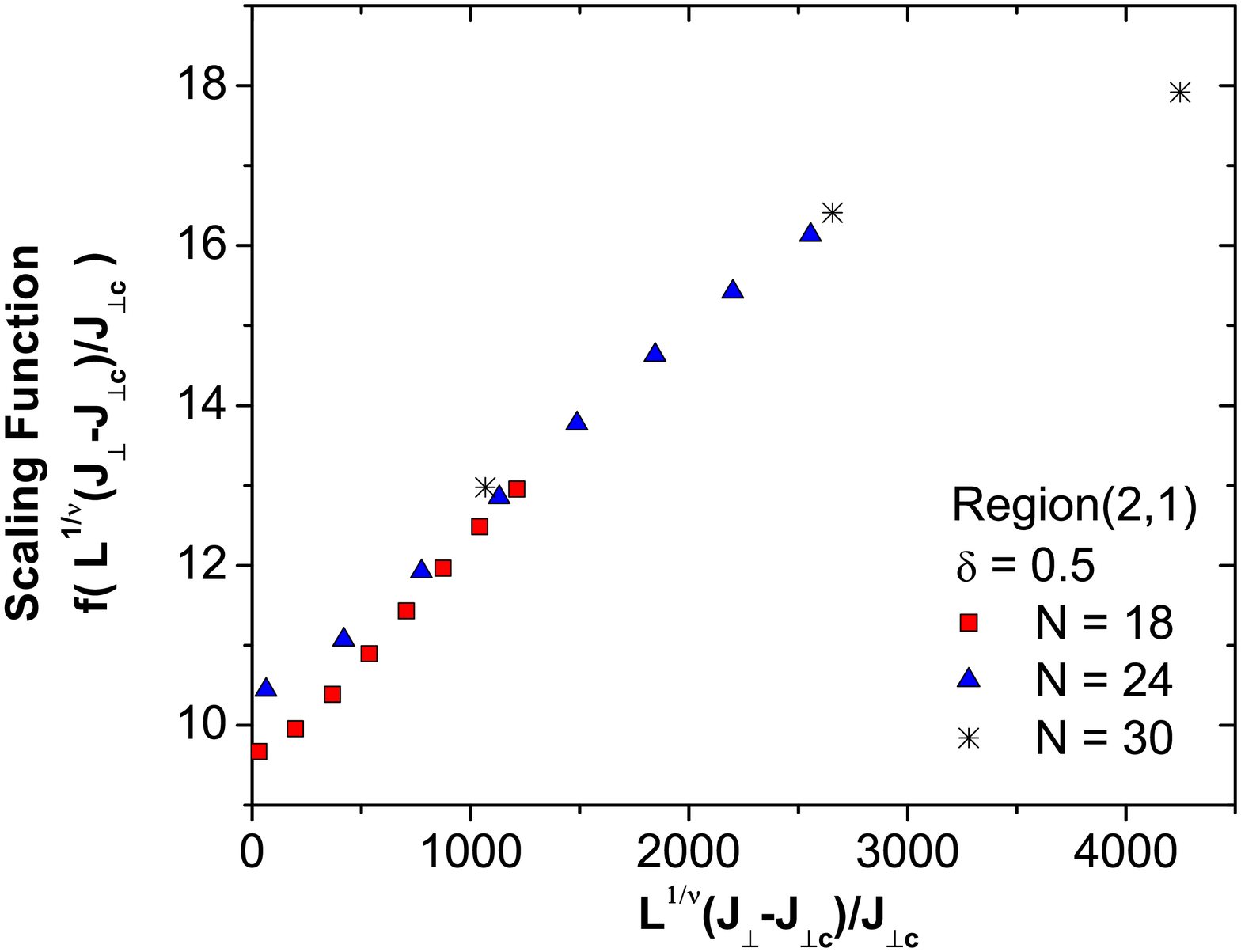}\hfil
\caption{(Color online) The collapsed scaling functions for the staggered
three-leg ladder with $\delta = 0.5$ in Region (1,2) (left panel) and
 Region (2,1) (right panel), calculated for $N = 12$ through 30 with optimized parameters
$\nu_{(1,2)} \approx 0.863$, $\nu_{(2,1)} \approx 0.385$ and $J_{\perp c} \approx 1.08$.
\label{scalingfx_3leg_collapse} }
\end{figure*}

%
%xxxxxxxxxxxxxxxxxxxxxxxxxxxxxxxxxxxxxxxxxxxxxxxxxxxxxxxxxxxxxxxxxxxxxxxxxxxxxx
The ground state energies were calculated by the ED method for the three-legged
ladder with N = 24 total number of spins. The columnar pattern was found to
have the consistently lower energies throughout the parameter space, in
agreement with the previous results from the bond mean-field
theory.\cite{Chitov08} The ED results for the energy gaps, given in
Fig.~\ref{gap_3leg} for N = 24, show that the columnar configuration is always
gapped, and the staggered pattern approaches gaplessness at a critical value of
rung coupling $J_{\perp c}$. Similar to the two-legged case, the appearance of
criticality in the three-legged ladder depends on the dimerization pattern. The
phase diagram of the staggered ladder is given in Fig.~\ref{Crit3L}. The ground
states of these ladders have been found to be well described by short-range
valence bond solid (VBS) states. A particular VBS state is denoted according to
the number of valence bonds formed with contiguous states; i.e., ($m,n$)-VBS
with $m+n = 3S$. A transition from a (1,2)-(VBS) phase to a (2,1)-VBS phase is
understood to occur on crossing the phase boundary.\cite{Almeida}
%
%xxxxxxxxxxxxxxxxxxxxxxxxxxxxxxxxxxxxxxxxxxxxxxxxxxxxxxxxxxxxxxxxxxxxxxxxxxxxxx
%xxxxxxxxxxxxxxxxxxxxxxxxxxxxxxxxxxxxxxxxxxxxxxxxxxxxxxxxxxxxxxxxxxxxxxxxxxxxxx
%
The finite size data for the energy gap was used to extract the critical indices
near the transition point. This was done using a critical scaling analysis similar
to that used to study the two-legged case, with the critical scaling relation given
by Eq.~\eqref{scaling1}. Data for laddders with $N = 12\mbox{ to } 30$ were used.
The scaling exponents on each side of the critical point,
denoted as Region (1,2) ($g < 0$) and Region (2,1) ($g > 0$) were found to be different,
with separate values $\nu_{(1,2)}$ and $\nu_{(2,1)}$. Collapse of the scaling function for
$\delta = 0.5$ (cf. Fig.~\ref{scalingfx_3leg_collapse}) yields $\nu_{(1,2)} \approx 0.863$,
$\nu_{(2,1)} \approx 0.385$ and $J_{\perp c} \approx 1.08$. For $\delta = 1$, the fitting
gives $\nu_{(1,2)} \approx 0.789$, $\nu_{(2,1)} \approx 0.363$ and $J_{\perp c} \approx 1.65$.
As opposed to the known behavior for large values of rung coupling $J_{\perp}$ of the gap
of the two-legged ladder, which increases as $\Delta \propto J_{\perp}$, the gap of the
three-legged ladder quickly saturates in both dimerization patterns
(cf. Fig.~\ref{gap_3l_long}). This fact limits the width of the critical region where
the finite-size scaling analysis is valid.
%
%xxxxxxxxxxxxxxxxxxxxxxxxxxxxxxxxxxxxxxxxxxxxxxxxxxxxxxxxxxxxxxxxxxxxxxxxxxxxxx
%xxxxxxxxxxxxxxxxxxxxxxxxxxxxxxxxxxxxxxxxxxxxxxxxxxxxxxxxxxxxxxxxxxxxxxxxxxxxxx
%
%xxxxxxxxxxxxxxxxxxxxxxxxxxxxxxxxxxxxxxxxxxxxxxxxxxxxxxxxxxxxxxxxxxxxxxxxxxxxxx
%xxxxxxxxxxxxxxxxxxxxxxxxxxxxxxxxxxxxxxxxxxxxxxxxxxxxxxxxxxxxxxxxxxxxxxxxxxxxxx
%
%xxxxxxxxxxxxxxxxxxxxxxxxxxxxxxxxxxxxxxxxxxxxxxxxxxxxxxxxxxxxxxxxxxxxxxxxxxxxxx
\begin{figure}[]
\includegraphics[width=8.0cm]{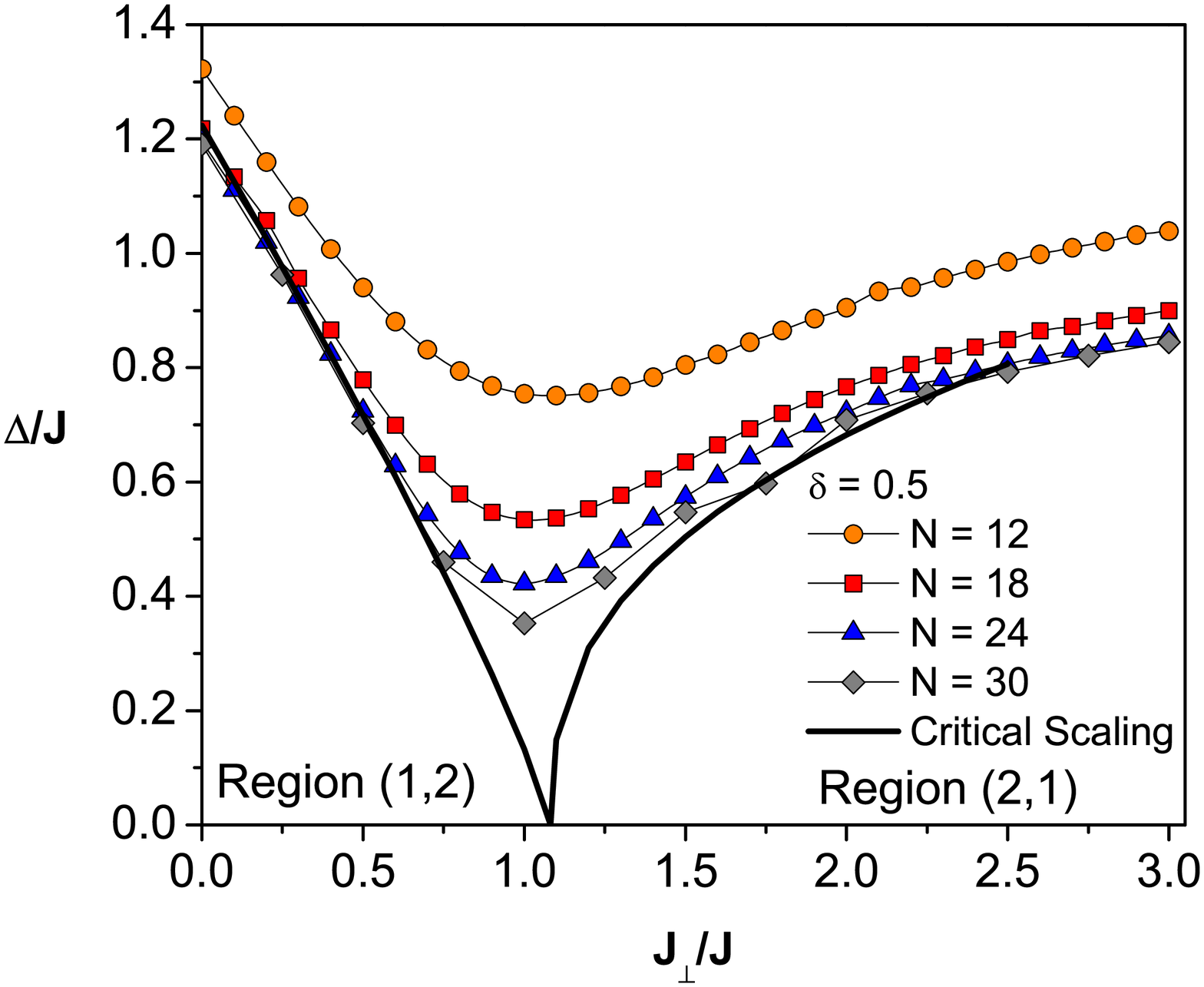}\hfil
\caption{(Color online) The ED data for the ground state energy gap with $\delta = 0.5$
for number of spins $N = 12$ to $N = 30$, along with the fit using optimized parameters
$\nu_{(1,2)} = 0.863$ in Region (1,2), $\nu_{(2,1)} = 0.385$ in Region (2,1) and
$J_{\perp c} = 1.08$. \label{3leg_stag} }
\end{figure}
%xxxxxxxxxxxxxxxxxxxxxxxxxxxxxxxxxxxxxxxxxxxxxxxxxxxxxxxxxxxxxxxxxxxxxxxxxxxxxx
%
%These results are summarized in Table~\ref{allindices_3l}.
%xxxxxxxxxxxxxxxxxxxxxxxxxxxxxxxxxxxxxxxxxxxxxxxxxxxxxxxxxxxxxxxxxxxxxxxxxxxxxx
%\begin{table}
%\begin{center}
%\caption{Critical scaling indices of the three-leg ladder.
%\label{allindices_3l} }
%\begin{tabular}{|cccc|} \hline \hline
%$\delta$ & $\nu_{(1,2)}$ & $\nu_{(2,1)}$ & $J_{\perp c}/J$\\ \hline \hline
%0.5 & 0.863 & 0.385 & 1.08 \\ \hline
%1   & 0.789 & 0.363 & 1.65\\ \hline
%\end{tabular}
%\end{center}
%\end{table}
%xxxxxxxxxxxxxxxxxxxxxxxxxxxxxxxxxxxxxxxxxxxxxxxxxxxxxxxxxxxxxxxxxxxxxxxxxxxxxx
%xxxxxxxxxxxxxxxxxxxxxxxxxxxxxxxxxxxxxxxxxxxxxxxxxxxxxxxxxxxxxxxxxxxxxxxxxxxxxx
%
\subsection{String Order Parameter}
\begin{figure}[hb!]
\includegraphics[width=8.0cm]{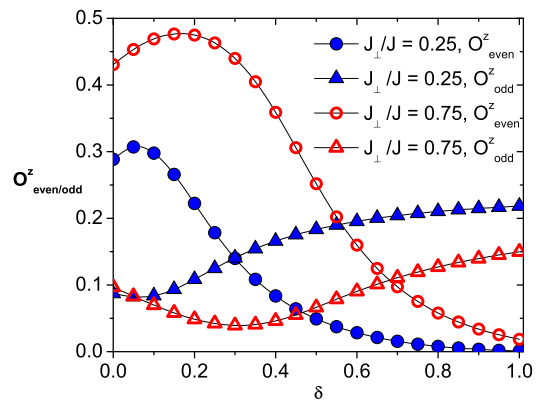}\hfil
\caption{(Color online) The odd and even string order parameters of the
staggered two-leg ladder with $J_\perp /J = 0.25$ and  $J_\perp /J = 0.75$ as
functions of dimerization $\delta$. The ED data for ladders with N = 24 total
number of spins is shown. \label{SOP_3leg_stag} }
\end{figure}

%
%xxxxxxxxxxxxxxxxxxxxxxxxxxxxxxxxxxxxxxxxxxxxxxxxxxxxxxxxxxxxxxxxxxxxxxxxxxxxxx
%
The generalized SOP for three-leg ladders is given by Eq.\eqref{stringorder} where
$S_\mathrm{rung}^z(n) \equiv S_1^z(n)+ S_2^z(n) + S_3^z(n)$ and
$S_\mathrm{diag}^z(n) \equiv S_1^z(n)+ S_2^z(n+1) + S_3^z(n+2)$. The SOP were calculated
for N = 24 for both the staggered and columnar configurations. Results for smaller systems
(N = 12-18) give unreliable results due to the very small number of rungs in the sum.
For this reason, a critical scaling analysis can not be performed for an estimate of
$\beta$ until much larger system sizes
(N = 30-36) are reached. Similar to the two-legged case, the columnar pattern was found
to possess even string order throughout the parameter space, with no critical features.
The staggered phase of the ladder possesses even string order associated with the (1,2)-VBS
phase for $0 < \delta < \delta_{c}$, which vanishes upon entering the (2,1)-VBS phase,
accompanied by emerging odd string order. Representative results for the SOP in the
staggered ladder are given in Fig. \ref{SOP_3leg_stag}. However, we note that a residual
odd string order persists in the (1,2)-VBS region of the parameter space. We believe
this effect is due to the very small size of the system, and should disappear if larger
system sizes are assessed.
%
%xxxxxxxxxxxxxxxxxxxxxxxxxxxxxxxxxxxxxxxxxxxxxxxxxxxxxxxxxxxxxxxxxxxxxxxxxxxxxx
%xxxxxxxxxxxxxxxxxxxxxxxxxxxxxxxxxxxxxxxxxxxxxxxxxxxxxxxxxxxxxxxxxxxxxxxxxxxxxx
%
%xxxxxxxxxxxxxxxxxxxxxxxxxxxxxxxxxxxxxxxxxxxxxxxxxxxxxxxxxxxxxxxxxxxxxxxxxxxxxx
%
\section{Conclusions}
%
%xxxxxxxxxxxxxxxxxxxxxxxxxxxxxxxxxxxxxxxxxxxxxxxxxxxxxxxxxxxxxxxxxxxxxxxxxxxxxx
%
We analyze the dimerized antiferromagnetic spin-$\frac12$ two- and three-leg
ladders. In agreement with the earlier mean-field predictions, \cite{Chitov08}
the present ED results demonstrate that the columnar phases have lower ground
state energies than the staggered phases. Thus, if no mechanism for locking
into a particular dimerization pattern is provided, a dimerized ladder would
order in the columnar phase. In this paper we analyze the two possible
dimerization patterns on the same footing, as intrinsic given properties of the
ladders without providing specific mechanisms responsible for a particular
dimerization.

The existence or absence of the quantum critical point is dependent on the
dimerization pattern. The gapped phases, which cannot be distinguished by the
conventional Landau long-range order parameter, have been characterized using
the non-local SOPs, which are non-zero in one phase and vanish in the other.
Here, for the first time, we systematically identify and numerically calculate
all SOPs for the both critical (staggered) and non-critical (columnar) phases
in the two-dimensional space of model's couplings ($\delta, J_\perp /J$). In
addition we analyze the critical parameters near quantum criticality.

For both 2- and 3-leg ladders the columnar pattern was found to be always
gapped and to possess the non-zero even SOP which is essentially featureless.
The finite-size ED data suggest that the odd SOP vanishes in the limit $N \to
\infty$ in the columnar phase for both types of ladders considered.

The finite size data for the energy gap were used to extract the critical
indices in the vicinity of the line of quantum phase transition in the
staggered case. The scaling behavior of the three-leg ladder was found to be
different on each side of the critical point. These results give for $\delta =
0.5$ values of $\nu_{(1,2)} \approx 0.863$ and $\nu_{(2,1)} \approx 0.385$.

From the critical scaling analysis of the ED data for the ladder with
two legs we find the critical indices of the correlation length $\nu \approx 0.755$,
of the odd order parameter $\beta_\mathrm{odd} \approx 0.2$ and of the even order
parameter of $\beta_\mathrm{even} \approx 0.02$. The rest of
the critical indices can be determined by the scaling relations.

Most likely, the smallness of $\beta_\mathrm{even}$ and large
difference between $\beta_\mathrm{even}$ and $\beta_\mathrm{odd}$
are due to stronger finite-size artifacts in $O_\mathrm{even}^{z}$.
The non-monotonous behavior of $O_\mathrm{even}^{z}$, clearly seen in
Fig.~\ref{SOP_2leg_stag}, also makes the finite-size scaling analysis
less reliable.

According to the predictions of
Refs.~[\onlinecite{Delgado96,Delgado98,Nersesyan00}] the critical behavior of
the two-leg ladder along the critical line $J_{\perp c}(\delta)$ is the same as
near the integrable point $(J_\perp /J=2, \delta=1)$.\cite{KimThanks} Since
along the line $\delta=1$ the ladder (\ref{Ham}) reduces to a dimerized
Heisenberg chain, the above assumption implies a critical behavior with $\nu
=2/3$ and  $\beta=1/12$ modified by multiplicative logarithmic corrections due
to the marginally irrelevant term in the effective Hamiltonian. Our values $\nu
\approx 0.75$ and  $\beta \approx 0.2$ (including the case $\delta =1$) are
comparable with these predictions, especially taking into account: \textit{(i)}
natural limitations on the precision for the critical indices due to the small
system sizes which can be reached by ED; \textit{(ii)} fitting the critical
indices without logarithmic corrections exaggerates their values (see, e.g.,
[\onlinecite{Okamoto86,ChainFits}]). The critical behavior of the ladders in
the vicinity (along) of the critical lines clearly warrants a detailed
analytical study. We plan to address this issue in our future work.
%
%xxxxxxxxxxxxxxxxxxxxxxxxxxxxxxxxxxxxxxxxxxxxxxxxxxxxxxxxxxxxxxxxxxxxxxxxxxxxxx
\begin{acknowledgments}
The authors thank E.H. Kim for
fruitful discussions. S.J.G. would also like to thank C. Gros, H. Merz and
A. Langille. Support from NSERC (Canada), Laurentian University Research Fund
(LURF), Laurentian University and the Shared Hierarchical Academic Research
Computing Network (SHARCNET) is gratefully acknowledged.
\end{acknowledgments}
%xxxxxxxxxxxxxxxxxxxxxxxxxxxxxxxxxxxxxxxxxxxxxxxxxxxxxxxxxxxxxxxxxxxxxxxxxxxxxx
%
%xxxxxxxxxxxxxxxxxxxxxxxxxxxxxxxxxxxxxxxxxxxxxxxxxxxxxxxxxxxxxxxxxxxxxxxxxxxxxx
% REFERENCES
%xxxxxxxxxxxxxxxxxxxxxxxxxxxxxxxxxxxxxxxxxxxxxxxxxxxxxxxxxxxxxxxxxxxxxxxxxxxxxx
%

%
%xxxxxxxxxxxxxxxxxxxxxxxxxxxxxxxxxxxxxxxxxxxxxxxxxxxxxxxxxxxxxxxxxxxxxxxxxxxxxx
%
\end{document}